\documentclass[reprint,amsmath,amssymb,aps,prb,superscriptaddress,floatfix,nofootinbib,nobibnotes]{revtex4-2}

\pdfoutput=1
\usepackage{graphicx}
\usepackage{dcolumn}
\usepackage{bm}
\usepackage{braket}
\usepackage[mathlines]{lineno}
\usepackage{stackengine}
\usepackage{verbatim}
\usepackage{graphics}
\usepackage[colorlinks]{hyperref}
\hypersetup{linkcolor = blue, citecolor = blue}
\usepackage{xcolor}
\usepackage{appendix}
\usepackage{soul}
\usepackage{ulem} 
\usepackage{multirow}
\usepackage{amsmath}
\usepackage{tikz}
\usepackage{amssymb}
\usepackage{nicefrac}
\usepackage{amsmath}
\usepackage{dsfont}
\usepackage{booktabs}

\begin{document}

	\title{Floquet physics from quantized light-matter interaction: geometric phases, gauge consistency, and entanglement}

\author{Beatriz P\'erez-Gonz\'alez}
\email{beatriz.perez.gonzalez@uni-a.de}
\affiliation{University of Augsburg$,$ Institute of Physics$,$ Universit\"atsstra\ss e 1 (Physik Nord)$,$ 86159 Augsburg, Germany}

\author{Sigmund Kohler}
\affiliation{Quantum Advanced Research Center (QuARC), CSIC, 28049 Madrid, Spain}
\affiliation{Instituto de Ciencia de Materiales de Madrid (ICMM), CSIC, C/Sor Juana In\'es de la Cruz 3, 28049 Madrid, Spain}

\author{ M\'onica Benito}
\affiliation{University of Augsburg$,$ Institute of Physics$,$ Universit\"atsstra\ss e 1 (Physik Nord)$,$ 86159 Augsburg, Germany}
\affiliation{Center for Advanced Analytics and Predictive Sciences, University of Augsburg, Augsburg, Germany}

\date{\today}

\begin{abstract}
	Floquet engineering and cavity quantum electrodynamics represent two complementary descriptions of light-matter interaction, yet their precise connection beyond the small-coupling regime remains subtle. Here, by representing the cavity field in a large-photon-number phase basis, we show that the time-dependent Schr\"odinger equation of a classically driven system emerges directly from the fully quantized problem, without replacing the field by a classical trajectory or assuming a mean-field decoupling. This establishes a one-to-one correspondence between quasienergies and photon-sector-shifted quantum energies, while providing a direct quantum interpretation of the Floquet mean energy and Anandan phase. Because gauge invariance is exactly preserved at the level of the quantum Hamiltonian, the quantum-to-classical correspondence remains valid beyond perturbative and small-coupling regimes. We further show that light--matter entanglement is encoded in matter--harmonic correlations of the corresponding Floquet mode, thus identifying a semiclassical limit in which Floquet physics emerges without requiring the quantum field itself to approach a coherent classical state.
\end{abstract}

\maketitle

\textbf{\textit{Introduction.}} Electromagnetic control of matter underpins modern quantum science \cite{Bloch2022}. Classical periodic drives enable coherent manipulation \cite{Silveri_2017}, dynamical band engineering \cite{takashi2019, Rudner2020, Bukov04032015, PhysRevB.90.125143, PhysRevLett.112.156801}, and phases without static counterparts \cite{ PhysRevX.3.031005, PRXQuantum.3.040312, GomezLeon2024anomalousfloquet, PhysRevB.82.235114, Quelle_2017}, while quantized fields store and transport quantum information \cite{PhysRevA.75.032329, Sillanpaa2007, Majer2007, Dijkema2025, xue2026controllableinteractionphotonsdistant, PhysRevX.12.021026, Guido2026, PRXQuantum.5.020339}, entangle with matter \cite{Blinov2004, PhysRevLett.131.023601, PhysRevLett.109.240501, PhysRevB.106.155113, Chirella2025}, and reshape quantum-material phases \cite{PhysRevB.99.235156, Dmytruk2022, PhysRevB.110.L121101, bretscher2026, Schlawin2022}. A central question is therefore how a semiclassical driven-system description emerges from a fully quantized light-matter theory, beyond reproducing qualitatively similar phenomena \cite{Perez-Gonzalez2024, PerezGonzalez2025lightmatter, PhysRevResearch.2.033033, Eckhardt2022}.

Coherent states \cite{Glauber1963Coherent, PhysRevD.32.400, PhysRevA.41.2645} are often viewed as the bridge between quantum and classical radiation because their expectation values approach classical waveforms at large photon number \cite{ Ashhab_2017, PhysRevA.79.032328, PhysRevA.95.062321, Peri2024beyondadiabatic, Coleman_24, PhysRevA.110.063707, Engelhardt2024, PhysRevResearch.6.013116}. Yet this concerns only the field sector and does not establish any equivalence between the corresponding light-matter problems. A complete mapping must instead be formulated at the Hamiltonian level. Such mappings can be obtained by displacing the photonic ground state \cite{PhysRevLett.129.183603, twyeffort2026quantumcorrectedfloquetdynamicsbridging} or from high-frequency expansions of effective Hamiltonians \cite{PhysRevResearch.2.033033, sueiro2025floquettheorylatticeelectrons, PhysRevB.105.165121}, within a restricted validity regime for both cases. These approaches ultimately invoke a factorization of light and matter, leaving unresolved whether a Floquet limit can emerge in the presence of genuine light–matter entanglement.

Further limitations concern gauge consistency and the quantum interpretation of Floquet-state properties. First, existing mappings generally ignore gauge choice, although it determines the form of the quantum Hamiltonian and therefore its semiclassical limit. A consistent mapping must track Hamiltonians, observables, and physical quantities across gauges \cite{PhysRevB.105.165121, PhysRevA.107.013722, DiStefano2019, PhysRevB.101.205140}. This requires models free from approximations that can break gauge invariance or spuriously restrict validity. Second, existing approaches only recover a Floquet Hamiltonian and its spectral properties, while a complete quantum correspondence for state properties like the mean energy, geometric phases, and possible light-matter entanglement has been elusive so far.

In this work, we bridge quantized light–matter interactions and Floquet physics through a phase representation of the photonic field \cite{PhysRevA.50.3505, PhysRevA.39.1665}. In the large-photon-number (Floquet) limit, the quantum Hamiltonian reduces to the Floquet operator without assuming a coherent-state distribution or imposing a mean-field separation of light and matter. We can then provide fully quantum interpretations of the Floquet quasienergies, mean energies, and Anandan phases, as well as the time-dependent and Sambe-space formulation of the Floquet problem. The framework also clarifies the role of gauge choice and shows that light-matter entanglement need not vanish in the semiclassical limit: the correlation structure of the quantum light–matter eigenstate is retained in the Floquet limit as non-factorization between matter spinors and Fourier harmonics in Sambe space. Lastly, we show that the photonic distribution of quantum eigenstates in the Floquet limit is far from that of a coherent state. Altogether, our results provide a unified framework for the quantum-to-classical correspondence in light--matter systems, with relevance to coherent control, engineered phases of matter, and light-enabled quantum technologies.\\

\textbf{\textit{Quantum and semiclassical Rabi Hamiltonians.}} As a workhorse model, we introduce the quantum Rabi model (QRM), describing a two-level system (TLS) coupled to a quantized electromagnetic field,
which in dipole gauge takes the form

\begin{equation}
	\hat{H}_{\text{QRM}}^{(\text{D})} = \Omega \hat{a}^\dagger \hat{a} + \hat{H}_{\mathrm{mat}} + ig\left(\hat{a} - \hat{a}^\dagger \right) \hat{Z} +  \frac{g^2}{\Omega} \cdot \openone
	\label{eq:quantum_dipole}
\end{equation}

where $g$ is the light-matter coupling strength in the dipole gauge, $\Omega$ is the cavity frequency, $\hat{a},\hat{a}^\dagger$ are the annihilation and creation operators of the field, and  $\hat{H}_\text{mat},\, \hat{Z}$ represent the TLS unperturbed Hamiltonian and coupling operator, respectively.
The final term is the dipole self-energy required for gauge consistency \cite{Olesya2021}, and cancels the coupling-dependent global shift generated by the linear interaction.. Its eigenvalue equation reads $\hat{H}_{\text{QRM}}^{(\text{D})}\ket{\varphi_\beta} = E_\beta \ket{\varphi_\beta}$, with eigenenergies $E_\beta$ and eigenstates $\ket{\varphi_\beta}$ ($\beta\in\mathbb N_0\}$), 
owing to the infinite dimension of the photonic Hilbert space. 

For comparison, let us consider the semiclassical Rabi model (SCRM) describing a 

TLS coupled to a linearly polarized driving field, 

\begin{equation}
	\hat{H}_{\mathrm{SCRM}}(t) = \hat{H}_{\text{mat}} + A \sin(\omega t) \, \hat{Z},
	\label{eq:SCRM}
\end{equation}

where $A$ is the driving amplitude. The dynamics of a periodically-driven system $i\partial_t \ket{\Psi(t)}=\hat{H}_{\rm SCRM}(t)\ket{\Psi(t)}$, with $\hat{H}_{\rm SCRM}(t)=\hat{H}_{\rm SCRM}(t+T)$, where $T$ is the driving period, can be treated using Floquet theory \cite{PhysRev.138.B979,PhysRevA.7.2203, Eckardt_2015, Bukov04032015}. Owing to discrete time-translation symmetry, the solutions to the Schr\"odinger equation can be written as
$\ket{\Psi_{\alpha}(t)} = e^{-i\varepsilon_{\alpha} t}\ket{\Phi_{\alpha}(t)}$, 
where $\ket{\Phi_{\alpha}(t)}=\ket{\Phi_{\alpha}(t+T)}$ are Floquet modes and $\varepsilon_{\alpha}$ are quasienergies ($\alpha=0,\dots,\dim(\mathcal{H}_{\mathrm{mat}}) - 1$). Since $\varepsilon_\alpha$ is defined as a phase factor, all $\varepsilon_{\alpha,k} = \varepsilon_\alpha + k\omega$ ($k\in\mathbb{Z}$) correspond to the same physical Floquet state. Hence, the quasienergy spectrum consists of infinitely many shifted copies of the fundamental set $\{ \varepsilon_\alpha \}$. Substituting the ansatz for $\ket{\Psi_\alpha (t)}$ into the Schr\"odinger equation yields an eigenvalue equation for the Floquet modes

\begin{equation}
	\left[\hat{H}(t)-i\partial_t\right]\ket{\Phi_{\alpha}(t)}=\varepsilon_{\alpha}\ket{\Phi_{\alpha}(t)},
	\label{eq:floquet_operator_eq}
\end{equation}

from which we define the Floquet operator $\hat{H}(t)-i\partial_t$. Its eigenvalues, the quasienergies, play the role of conserved quantities for driven systems. For convenience, we shall refer to the Rabi model in Eq. \eqref{eq:quantum_dipole} and Eq. \eqref{eq:SCRM}, as (fully) quantum / semi-classical, respectively, in accordance with the nature of the bosonic mode.\\

\textit{\textbf{Quantum-to-classical mapping.}} A common route to connect the two Hamiltonians is to replace the photonic operators by their expectation values, assuming a coherent cavity field, $\beta(t) = \langle \hat{a}(t)\rangle_{\mathrm{coh}}=\beta e^{-i\Omega t}$ with $\beta=\sqrt{\langle \hat{a}^\dagger \hat{a}\rangle}$. A more systematic approach applies the displacement operator $\hat{D}[\beta(t)]=\exp[\beta(t)\hat{a}-\beta^*(t)\hat{a}^\dagger]$ to Eq.~\eqref{eq:quantum_dipole}, later requiring the factorization of light and matter subspaces with $2g\beta \equiv A$  ($\beta \rightarrow \infty$, $g\rightarrow 0$) \cite{PhysRevLett.129.183603}.  
In both cases, the semiclassical limit ultimately relies on a mean-field decoupling of subsystems, implicitly assuming that the two remain separable.

We now show that the Floquet description emerges directly from the fully quantized light-matter Hamiltonian, without invoking coherent states, and keeping the full hybridization structure between subsystems. For this, we introduce the following basis for the cavity photons \cite{PhysRevA.50.3505, PhysRevA.39.1665},

\begin{equation}
	\ket{\theta} = \sum_{n = 0}^\infty e^{-i(n - n_0)\theta}\ket{n},
	\label{eq:theta_definition}
\end{equation}

as a linear combination of Fock states $\{\ket{n}\}$ ($n \in \mathds{Z}_{\geq 0}$), $\theta$ is a continuous and $2\pi-$periodic variable, and $n_0$ represents a reference photon number. In this representation, $(\hat{a}^\dagger \hat{a})_\theta = n_0 -i\partial_\theta$. Because the Fock ladder is bounded from below, these states do not form an exact Fourier basis. They become a true phase representation asymptotically as $n_0\rightarrow\infty$, when $\langle \theta |\theta' \rangle = 2\pi \delta_{2\pi}(\theta-\theta') $ (End Matter, Appendix \ref{app:theta_basis}). In this limit, the linearized light--matter coupling becomes local in $\theta$, $ i(\hat{a} - \hat{a}^\dagger)_\theta = 2\sqrt{n_0}\sin(\theta)$. Now, by further requiring the Floquet limit to be $2g\sqrt{n_0} \equiv A$ ($n_0 \rightarrow \infty$), the eigenvalue equation for the quantum Hamiltonian takes the form,

\begin{equation}
	\left[\hat{h}_{\rm QRM}^{\rm (D)}(\theta ) - i\Omega \partial_\theta\right]\ket{\varphi_\beta (\theta)} = (E_\beta - \Omega n_0) \ket{\varphi_\beta (\theta)} ,
	\label{eq:schr_theta_eq}
\end{equation}

where $\hat{h}_{\rm QRM}^{\rm (D)}(\theta )= \hat{H}_{\text{mat}}  + A \sin(\theta)\,\hat{Z}$. This exactly corresponds to the eigenvalue equation for the Floquet-driven system of Eq.~\eqref{eq:floquet_operator_eq}, with $\theta$ playing the role of a rescaled time $\theta \equiv \Omega t$. Note that the self-dipole energy term vanishes in this limit, as $g^2/\Omega = A^2 / 4n_0\Omega \rightarrow 0$. The $\theta-$resolved $\hat{h}^{\rm (D)}_{\rm QRM}(\theta)$ has dimensions of $\mathrm{dim}(\mathcal{H}_{\rm mat}) \times \mathrm{dim}(\mathcal{H}_{\rm mat})$, while the photonic field appears as a parametric $\theta$ dependence.\\

\begin{table}
	\centering
	\caption{Corresponding quantities between the Floquet problem and the fully quantum theory.}
	\label{tab:example}
	\begin{tabular}{lr}
		\toprule[1.5pt]
		Quantum & Floquet \\
		\midrule
		$E_\beta-\Omega n_0$ & Quasienergies \\
		$\ket{\varphi_\beta(\theta)}$ & Floquet modes \\
		$n_0$ & Floquet--Brillouin zone \\
		$2g\sqrt{n_0}$ & Driving amplitude, $A$ \\
		$n_0-\langle\hat{a}^\dagger\hat{a}\rangle_\beta
		=\displaystyle\oint d\theta 
		\langle i\partial_\theta\rangle_\beta / 2\pi$
		& Anandan phase  \\
		$\begin{aligned}
			E_\beta -\Omega\langle\hat{a}^\dagger\hat{a}\rangle_\beta  =\oint d\theta 
			\langle
			\hat{h}_{\rm QRM}(\theta) \rangle_\beta / 2\pi
		\end{aligned}$
		& Mean energy\\
		\bottomrule[1.5pt]
	\end{tabular}
\end{table}

Comparison between Eqs.~\eqref{eq:floquet_operator_eq} and \eqref{eq:schr_theta_eq} gives a transparent physical meaning to distinct features of Floquet theory. Firstly, we can formally identify the quasienergies $\varepsilon_{\alpha}$ with the shift-corrected energies of $(E_\beta - n_0\Omega)$, as expected. Changing the reference occupation by an integer $n_0\rightarrow n_0 + p$ corresponds to a Floquet-zone shift of the quasienergies $\varepsilon_{\alpha} \rightarrow \varepsilon_{\alpha} + p\omega$, which gives an extra phase in $\ket{\Phi_{\alpha,p}(t)} = e^{-ip \omega t}\ket{\Phi_\alpha(t)}$ while leaving the full physical state $\ket{\Psi_\alpha(t)}$ unchanged. Hence, the modulo-$\omega$ replica structure of Floquet theory is the 
semiclassical remnant of the gauge freedom to choose $n_0$ in the quantum theory.  Secondly, the operator \(-i\partial_t\) in the Floquet eigenvalue equation [Eq. \eqref{eq:floquet_operator_eq}] emerges naturally from $\hat{n}_\theta = n_0 -i\partial_\theta $, with $\partial_\theta$ counting excitations on top of a background of $n_0$ photons.

Equation~\eqref{eq:schr_theta_eq} also clarifies the quantum origin of the geometric phase of the Floquet states.  The total phase picked by a Floquet state $\ket{\Psi_{\alpha}(t)}$ under a one-period evolution, $-\varepsilon_\alpha T$, can be separated into a  dynamical contribution, i.e., the mean-energy $ \overline{H}_\alpha$, and a geometric one, i.e., the Anandan-Floquet phase \cite{PhysRevLett.58.1593, PhysRevD.38.1863}, such that $\varepsilon_\alpha = \overline{H}_\alpha - (\omega/2\pi)\gamma^\text{F}_\alpha$, see Supplemental Material (SM) \ref{app:anandan_phase}. One can define an analogous quantum Anandan phase $\gamma_\beta^{\text{Q}}/2\pi \equiv n_0-\langle \hat{a}^\dagger \hat{a}\rangle_\beta$, and a quantum mean energy, $\overline{E}_\beta = E_\beta - \Omega \langle \hat{a}^\dagger \hat{a}\rangle_\beta$, for an eigenstate $\ket{\varphi_\beta(\theta)}$ of the quantum Hamiltonian. Table \ref{tab:example} summarizes the equivalences, with all expected values written as $\langle \hat{[\cdot]} \rangle_\beta = \langle \varphi_\beta \vert \hat{[\cdot]} \vert \varphi_\beta \rangle $. The quantum objects also satisfy $E_\beta  - \Omega n_0 =  \overline{E}_\beta - (\Omega/2\pi)\gamma^{\mathrm{Q}}_\beta $. Therefore, the dynamical phase of the Floquet state is replaced by the matter average energy in the quantum theory, while the geometric one is replaced by the averaged photonic excitation operator $i\partial_\theta$. \\

\begin{figure}[!t]
	\centering
	\includegraphics{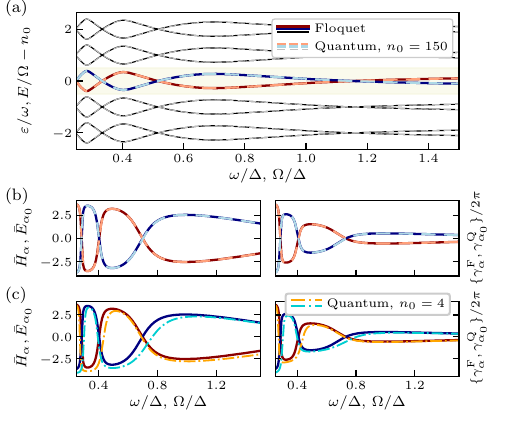}
	\caption{Semiclassical Floquet (solid) vs quantum (dashed lines): (a) quasienergies and energies, (b) mean energies (left panel), and Anandan phases (right panel), all for  $n_0 = 150$. (c) Mean energies (left panel) and Anandan phases (right panel), for $n_0 = 4$. For all plots, $A/\Delta=1.25$. In (a), the Floquet-Brillouin zone is shaded.}
	\label{fig:quasi_meane_anadan}
\end{figure}

We now particularize our results to $\hat H_{\mathrm{mat}}=(\Delta/2)\hat\sigma_x$ and $\hat Z=\hat\sigma_z$, where $\Delta$ is the tunneling amplitude. In the numerical calculations, we retain the dipole self-energy term, since its contribution $g^2/\Omega=A^2/(4n_0\Omega)$ for finite $n_0$ depends on the interplay of all parameters involved, and is required for a good agreement between energies and quasienergies. Figure~\ref{fig:quasi_meane_anadan}(a) compares the SCRM quasienergy spectra $\varepsilon_\alpha$ ($\alpha=0,1$)
with the quantum spectrum $E_{\beta}/\Omega - n_0$, as a function of $\omega/\Delta$ and $\Omega/\Delta$, respectively. For fixed $n_0$ and $A$, the scaling $2g\sqrt{n_0}=A$ ensures that the quantum eigenstates with index $\beta_0 = 2n_0 + \alpha$ reproduce the Floquet solution in the first Floquet-Brillouin zone, for arbitrary parameter choices. The two  relevant states ($\alpha = 0,1$) within the first Brillouin zone (shaded region) are ordered by ascending energy for the smallest $\omega, \Omega$ considered, and adiabatically followed in the parameter sweep, as indicated by the red and blue colors in the plot. The agreement between $\{ \gamma^{\mathrm{F}}_\alpha, \gamma^{\mathrm{Q}}_{\beta_0}\}$ and $\{ \overline{H}_\alpha, \overline{E}_{\beta_0} \}$ shown in panel \ref{fig:quasi_meane_anadan}(b) is excellent as well. For sufficiently large $n_0$, this correspondence extends over a broad energy window, as $\sqrt{n_0\pm k}\approx \sqrt{n_0}$, where $k$ labels neighboring photonic subspaces [Fig.~\ref{fig:quasi_meane_anadan}(a)]. Reducing $n_0$ progressively narrows the energy window over which the mapping remains valid, yet the quantum eigenstates $\beta_0$ can still reproduce the structure of the corresponding Floquet mode with high accuracy within a certain parameter regime. Surprisingly small photon occupations ($n_0 = 4$) already reproduce $\overline{H}_\alpha$ and $\gamma_\alpha$ for higher frequencies [Fig.~\ref{fig:quasi_meane_anadan}(c)]. Hence, a meaningful local quantum-Floquet correspondence can already emerge well before the global limit is reached (SM \ref{app:disagreement}).\\

\textbf{\textit{Connection to Sambe representation.}} Returning to Eq. \eqref{eq:schr_theta_eq}, we can leverage the $2\pi-$periodicity of $\theta$ to expand the quantum objects in Fourier harmonics, using $\hat{h}^{\rm (D)}_{\text{QRM}}(\theta) = \sum_k \hat{h}_{\text{QRM}}^{({\rm D}, k)} e^{-ik\theta}$,  and $\ket{\varphi_\alpha(\theta)}=\sum_k \ket{\varphi_{\alpha}^{(k)}} e^{-ik \theta}$, with Fourier coefficients $[\cdot]^{(k)}=\frac{1}{2\pi}\int_0^{2\pi} dt\, [\cdot](t)e^{+ik\theta}$. One arrives at

\begin{equation}
	\sum_l [\hat{h}_{\rm QRM}^{(\mathrm{D}, k - l)} - l\Omega\delta_{l,k}]\ket{\varphi_{\beta}^{(l)}}  = (E_\beta - \Omega n_0) \ket{\varphi_{\beta}^{(k)}}.
	\label{eq:fourier_theta_quantum}
\end{equation}

This can be directly connected to the Sambe-space version of Eq.~\eqref{eq:floquet_operator_eq} in Fourier harmonics, with the identification $\hat{H}_{\mathrm{SCRM}}(t)\leftrightarrow \hat{h}^{\rm (D)}_{\mathrm{QRM}}(\theta)$, after excluding the cavity free-energy term. Equation~\eqref{eq:fourier_theta_quantum} indicates that each $\hat{h}^{\rm (D,k)}_{\mathrm{QRM}}$ therefore maps onto the $k$th Floquet harmonic, making explicit the interpretation of the Sambe index as an effective photon-number index. Additionally, we can write $
\hat{h}^{\rm (D,k)}_{\mathrm{QRM}} = \langle n_0+k|
\bigl[\hat{H}^{\rm (D)}_{\rm QRM}-\Omega\hat a^\dagger\hat a\bigr]
|n_0\rangle$, such that the Fourier harmonics of the $\theta$-resolved Hamiltonian are precisely associated with $k$-photon exchange processes in the Floquet limit, with matrix elements depending only on the photon-number difference $k$, and not on the reference occupation $n_0$. The $\theta$ representation consequently establishes a direct correspondence not only between the quantum Hamiltonian and the Floquet operator through the $2\pi$-periodic phase coordinate, but also between their Fock-space and Sambe-space structures.\\

\textbf{\textit{The role of gauge choice.}} The framework is not limited to dipole-gauge Hamiltonians and applies equally in the Coulomb gauge, where light--matter coupling enters through a minimal-coupling phase \cite{Olesya2021,PhysRevResearch.3.023079, DiStefano2019}. For the case considered above, with $\hat{H}_{\text{mat}} = \Delta\hat{\sigma}_x/2$ and $\hat{Z} = \hat{\sigma}_z$, and using $\hat{U}_{\text{gauge}} = \mathrm{exp}[i(g/\Omega)(\hat{a} + \hat{a}^\dagger)\hat{\sigma}_z]$, the Coulomb-gauge Hamiltonian has the following form (SM \ref{app:gauge}),

\begin{equation}
	\hat{H}_{\text{QRM}}^{(\text{C})} = \Omega \hat{a}^\dagger \hat{a} \, + \,\frac{\Delta}{2}\left[ e^{-i\eta(\hat{a} + \hat{a}^\dagger)} \, \hat{\sigma}_+ + e^{i\eta(\hat{a} + \hat{a}^\dagger)} \, \hat{\sigma}_{-}\right].
	\label{eq:QRM_Coulomb}
\end{equation}

where $\eta = 2g/\Omega$ can be defined as the Coulomb-gauge light-matter coupling strength. The mapping to the Floquet limit can be also performed in this gauge, which leads to the $\theta$-dependent Peierls phase $\mathrm{exp}[\pm i (4g\sqrt{n_0}/\Omega) \cos(\theta)]$. Again, by identifying $2g\sqrt{n_0} \equiv A$ and $\{\theta \equiv \omega t,\Omega \rightarrow \omega \}$, one arrives at the Coulomb-frame driven Hamiltonian,

which is precisely the Hamiltonian obtained by rotating the semiclassical Rabi Hamiltonian in Eq. \eqref{eq:SCRM} with $\hat{U}_{\mathrm{driv}}(t) = \mathrm{exp}\left[-i A  \hat{\sigma}_z \int \sin(\omega t)\right]$, that we shall refer to as $\hat{H}^{\prime}_{\rm SCRM}(t)$. Because gauge invariance is exactly preserved, we can identify $\hat{U}_{\rm gauge}$ to this rotating-frame transformation upon $\hat{a}+\hat{a}^\dagger \to 2\sqrt{n_0}\cos\theta$ in the Floquet limit.\\

Because the dipole- and Coulomb-gauge Hamiltonians are exactly unitarily related, gauge invariance ensures that their quantum spectra converge to the same Floquet quasienergies throughout parameter space. This correspondence may fail for approximate Rabi models not connected by such a transformation, spuriously restricting the quantum-to-classical crossover to regimes of approximate gauge invariance. Although the mapping is commonly studied using linearized quantum Rabi Hamiltonians, both gauges yield linear light-matter couplings after specific approximations. In the Coulomb gauge, expanding the Peierls phase to linear order in $g$  gives a model resembling Eq.~\eqref{eq:SCRM} under $2g_{\rm C}\sqrt{n_0}\equiv A$ ($g_{\rm C} = \Delta g /\Omega$), up to a matter-space rotation. To the same order in $g$, neglecting the dipole self-energy in the dipole gauge yields two Hamiltonians whose semiclassical limit differs only by a $\pi/2$ drive-phase shift. The gauge of such linearized models is often left implicit \cite{Ashhab_2017,PhysRevLett.129.183603}, obscuring whether the Coulomb-gauge diamagnetic term or dipole self-energy was neglected. Note that the microscopic description of the light-matter coupling also depends on the gauge. This ambiguity underlies debates such as the Dicke-model superradiant phase transition \cite{PhysRevLett.35.432,Keeling_2007,PhysRevLett.112.073601,PhysRevA.97.043820,PhysRevA.98.053819,PhysRevA.7.831,Stokes2019} and likewise affects the quantum-to-classical correspondence: inconsistent truncations yield mismatched quantum spectra and Floquet quasienergies outside their 
validity regimes. The conclusion is that linearized models recover the Floquet spectrum only for small $A$, equivalently small $g$, whereas 
the gauge-consistent approach in this work guarantees validity throughout parameter space. Agreement restricted to small coupling or high frequency therefore reflects approximate gauge invariance, not a fundamental limitation of the crossover (SM \ref{app:non-gi-models}).\\

\begin{figure}[t!]
	\centering
	\includegraphics{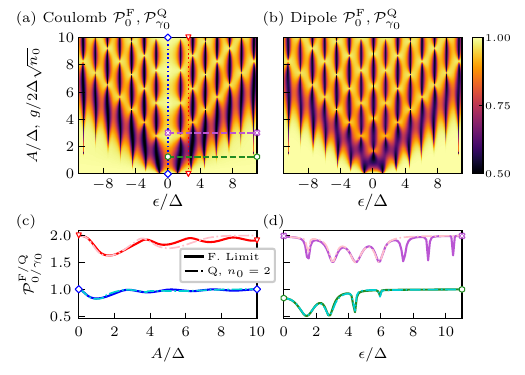}
	\caption{
		Floquet and quantum purity $\{\mathcal{P}^{\mathrm{F}}_{0}, \mathcal{P}^{\mathrm{Q}}_{\gamma_0} \}$ ($n_0 = 150$) in the (a) Coulomb, and (b) dipole gauge, as functions of $\epsilon/\Delta$ and $A/\Delta$. Dotted and dashed lines mark $\epsilon /\omega = 0, 5/2$ and $A/\Delta = 1.25, 3$. The lower panels (c), (d), show those cuts, comparing the Floquet result with the quantum
		model for $n_0=2$. The upper lines have been shifted for clarity. For all plots, $\Omega = 1.5\Delta$.}
	\label{fig:entanglement}
\end{figure}

\textbf{\textit{Light-matter entanglement}.} Another key facet of the quantum–to–classical crossover concerns the light--matter entanglement. It is often argued that in the semiclassical limit, light–matter entanglement should vanish, since genuine quantum correlations are absent \cite{PerezGonzalez2025lightmatter, PhysRevLett.129.183603}. Here we show to what extent light-matter entanglement is still traceable in the semiclassical limit, and that the crossover still exists even for parameter choices in which light and matter are maximally entangled. 
Given the fully quantum eigenstate $\ket{\varphi_{\beta_0}}$  of the quantum Hamiltonian, we quantify light-matter entanglement through the purity of $\mathcal{P}^{\text{Q}}_{\beta_0} = \mathrm{Tr}[(\rho_{\beta_0}^{\text{Q}})^2]$, where $\rho^{\text{Q}}_{\beta_0} = \mathrm{Tr}_\text{ph}(\ket{\varphi_{\beta_0}}\bra{\varphi_{\beta_0}})$. For our system, $1/2 \le \mathcal{P}^\text{Q}_{\beta_0} \le 1$, with $\mathcal{P}^\text{Q}_{\beta_0} = 1/2$ being maximally mixed subsystems. Remarkably, the quantum purity $\mathcal{P}^{\mathrm{Q}}_{\beta_0}$ maps directly onto the purity obtained by tracing over the harmonic degree of freedom in Sambe space as if they constituted a true dynamical degree of freedom of the driven system. For a Floquet mode
$\ket{\Phi_\alpha}\rangle=\sum_k\ket{\Phi_\alpha^{(k)}}\otimes\ket{k}$, with $\ket{k}\equiv e^{ik\omega t}$ spanning $\mathcal{T}$, this gives $\rho^{\mathrm{F}}_\alpha=\sum_{k\in\mathds{Z}}
\ket{\Phi_\alpha^{(k)}}\bra{\Phi_\alpha^{(k)}}$ and $\mathcal{P}^{\mathrm{F}}_\alpha = \mathrm{Tr}[(\rho^{\mathrm{F}}_\alpha)^2]$. In the Floquet limit, $\mathcal{P}^{\mathrm{F}}_\alpha$ and $\mathcal{P}^{\mathrm{Q}}_{\beta_0}$ coincide, for $\beta_{n_0} \equiv 2n_0 + \alpha$.
To identify further patterns and expand the scope of the comparison, we will also include a static energy splitting for the qubit $\hat{H}_\text{mat} = (\epsilon / 2) \hat{\sigma}_z + (\Delta/2)\hat{\sigma}_x$ to both the quantum and semiclassical description. For the Floquet purity, we diagonalize a sufficiently large representation of the Sambe Hamiltonian ($k_{\rm max} = 100$), and take the mode with the lowest energy out of the two. The results are shown in Fig. \ref{fig:entanglement}. Under the constraints of the Floquet limit, $\mathcal{P}^\text{Q}_{\beta_0}$ and $\mathcal{P}^\text{Q}_\alpha$ yield the exact same result \textit{when compared in the same gauge/frame}. Panel \ref{fig:entanglement}(a) [\ref{fig:entanglement}(b)] is obtained using $\hat{H}^{\prime}_{\rm SCRM}(t)$ and $\hat{H}^{\rm (C)}_{\rm QRM}$ ($\hat{H}_{\rm SCRM}(t)$ and $\hat{H}^{\rm (D)}_{\rm QRM}$). 
The main difference between frames appears at $\epsilon/\omega = 0$: in the Coulomb gauge (frame), the light-matter (spinor-harmonic) mixing is zero, whereas in the dipole gauge (frame), it is maximal. In the quantum case, we recall that the electromagnetic field in the dipole gauge is redefined as a mixture of light and photonic degrees of freedom $\hat{a}_D = \hat{U}^\dagger_{\text{gauge}}\, \hat{a} \,U_{\text{gauge}} = \hat{a} - i (g/\Omega)\hat{\sigma}_z $ \cite{PhysRevResearch.3.023079}. This redefinition is inherited in the Floquet system, with the same frame dependence as in the fully quantum theory, providing a clear example of the role of gauge choice in the crossover. In panels \ref{fig:entanglement}(c) and (d) we compare for different $\epsilon/\omega$ the agreement at low $n_0$, which again confirms that a local semiclassical limit is possible within certain parameter regimes. With this, we have established a correspondence between the harmonic-spinor correlation of a Floquet mode and the degree of light–matter entanglement in the hybrid eigenstates of the quantum model.\\

\begin{figure}[t]
	\centering
	\includegraphics[width=\columnwidth]{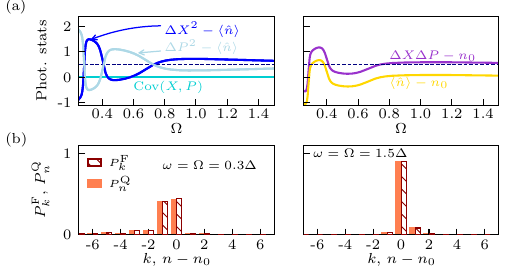}
	\caption{(a) Coulomb-gauge photon-statistics indicators versus $\Omega$, for fixed $A= 1.25\Delta$: $\Delta X^2-\langle\hat n\rangle$, $\Delta P^2-\langle\hat n\rangle$, $\mathrm{Cov}(X,P)$, $\Delta X\,\Delta P$, and $\langle\hat n\rangle-n_0$. (b) Quantum photon-number distribution \(P_n^{\mathrm Q}\) and Floquet harmonic weights \(P_k^{\mathrm F}\) at $\omega,\Omega=0.3\Delta$ and $1.5\Delta$, respectively. For all plots, $n_0 = 150$.}
	\label{fig:weights}
\end{figure}

\textbf{\textit{Photon statistics.}} One can additionally compute the statistics of the photonic distribution of $\ket{\varphi_{\beta_0}}$ in the Floquet limit, to show that they need not coincide with that of a coherent state. For fixed $A = 1.25\Delta$, Fig. \ref{fig:weights}(a) (left panel) shows the equal-and-opposite oscillations of the two quadrature variances $\Delta X^{2}$ and $\Delta P^{2}-\langle\hat n\rangle$, while the covariance $\mathrm{Cov}(X,P)$ remains zero. For $\Omega \gg \Delta$, $\Delta X^2 =\Delta P^2 = \langle \hat{n}\rangle + 1/2$, which corresponds to a Fock-state variance for occupation \(n\simeq\langle\hat n\rangle\). As both $\Delta X^2,\Delta P^2 \sim \langle \hat{n}\rangle$, they do not describe a coherent state. The product $\Delta X \Delta P$ also exhibits oscillations, correlated to those of $\langle \hat{n} \rangle$, as shown in the right panel. Panel \ref{fig:weights}(b) shows that the quantum photonic weights $P^{\mathrm Q}_{n - n_0} = \langle n - n_0 \vert \rho^{\mathrm{Q}}_{\gamma_0}\vert n - n_0\rangle$ for a given $A/\Omega$ reproduce the weights of Floquet harmonics $\mathcal{P}^{\mathrm F}_{k} = \langle k\vert \rho^{\mathrm{F}}_\gamma\vert k \rangle$ under the correspondence $n-n_{0}\leftrightarrow k$, as an additional consequence of the direct connection between the photon sectors of the quantum model with the harmonics of its Floquet counterpart.\\

\textbf{\textit{Conclusions.}} We have shown how Floquet theory emerges directly from a fully quantized light–matter Hamiltonian through the phase representation of the photonic field. Beyond the correspondence between quantum energies and quasienergies, our framework provides quantum counterparts of the Floquet mean energy and Anandan phase and identifies photon sectors with Fourier harmonics in Sambe space. The mapping is gauge consistent and, crucially, does not require coherent radiation or vanishing light–matter entanglement: the correlation structure of the entangled quantum state is encoded in matter–harmonic correlations of the corresponding Floquet mode. This establishes a broader notion of the quantum-to-classical crossover in which the effective dynamics becomes classical while the underlying field state need not. Our results therefore show that the emergence of Floquet physics from quantized light does not require the field itself to become classical. Future work will address generalized Rabi and Dicke models \cite{Duan_2022, rgz2-4m69}, the extension to open systems \cite{10.1063/5.0269753, keliri2026}, and connection to topological structures in quantum light-matter models \cite{kohler2026, PhysRevB.86.115318, GomezLeon2024anomalousfloquet, Perez-Gonzalez2024}.\\

\textbf{\textit{Acknowledgments.}}
We thank Marcus Kollar, Daniel Braak, \'Alvaro G\'omez-Le\'on, and Gloria Platero for fruitful discussions. 
BPG acknowledges a postdoctoral fellowship from the Alexander von Humboldt Foundation.
This work was supported by the Emmy Noether Programme of the German Research
Foundation (DFG) under grant no.\ BE 7683/1-1 and by the Spanish Ministry of Science,
Innovation, and Universities under grant nos.\ PID2023-149072NB-I00 and AIA2025-163435-C44.

	\bibliography{bibliography}

	\appendix

\section*{END MATTER}

\section{\texorpdfstring{$\vert \theta \rangle$}{theta} basis states and matrix elements of the fully quantum Hamiltonian \label{app:theta_basis}}

\subsection{General properties of the basis \texorpdfstring{$\{\ket{\theta}\}$}{\{theta basis\}}.} 

Let us first explore the properties of these newly defined states $\ket{\theta}$. We can easily show that 

\begin{equation}
	\int_0^{2 \pi}  \frac{d \theta}{2\pi} |\theta\rangle\langle\theta|  =  \sum_{n, m=0}^{\infty}|n\rangle\langle m| \int_0^{2 \pi} \frac{d \theta}{2 \pi} e^{i(m-n) \theta}
	=  \mathds{1}
\end{equation}

so the completeness relation is generally satisfied, without further assumptions. Now, for the overlap between two states $\ket{\theta}$ and $\ket{\theta^\prime}$ we obtain

\begin{eqnarray}
	\left\langle\theta \mid \theta^{\prime}\right\rangle= e^{in_0(\theta^\prime - \theta)}\sum_{n=0}^{\infty} e^{-i n\left(\theta^{\prime}-\theta\right)} .
\end{eqnarray}

As the sum is bounded from below with $n = 0$, this does not correspond to a $2\pi-$periodic delta distribution or Dirac comb,

\begin{align}
	\delta_{2\pi}(\theta - \theta^\prime) & \equiv \sum_m\delta(\theta-\theta^\prime - 2\pi m) \nonumber \\
	& = \frac{1}{2\pi}\sum_{m \in \mathds{Z}} e^{-im(\theta-\theta^\prime)},
\end{align}

which would act on any $2\pi-$periodic function $f(\theta) = f(\theta + 2\pi)$ as

\begin{equation}
	\int^{2\pi}_0 d\theta^\prime\,\delta_{2\pi}(\theta-\theta^\prime)f(\theta^\prime) = f(\theta).
	\label{eq:dirac_comb}
\end{equation}

\subsection{Matrix elements of photonic operators in the \texorpdfstring{$\theta$}{theta}-representation.} 

First, we transform the annihilation operator $\hat{a}$, using $\hat{a}|n\rangle = \sqrt{n}|n-1\rangle$. From this, and using the definition of $\ket{\theta}$ in Eq. \eqref{eq:theta_definition}, we find

\begin{equation}
	\langle \theta \vert \hat{a}\vert \theta^\prime \rangle = e^{-i\theta^\prime} \sum_{m = 0}^\infty \sqrt{m + 1} e^{-i(m-n_0)(\theta^\prime - \theta)},
	\label{eq:a_matrix_elements}
\end{equation}

where the non-locality resides in the $m-$dependent prefactor. For the number operator $ \hat{n} \equiv \hat{a}^{\dagger} \hat{a}$, we get 

\begin{equation}
	\hat{a}^{\dagger} \hat{a}|\theta^\prime \rangle = \sum_{n=0}^{\infty} n e^{-i (n - n_0) \theta^\prime}|n\rangle = (n_0 + i \partial_{\theta^\prime})|\theta^\prime\rangle,
\end{equation}

The relevant matrix elements are

\begin{equation}
	\langle \theta \vert \hat{a}^\dagger \hat{a} \vert \theta^\prime \rangle = \sum_{n = 0}^\infty n e^{-i(n - n_0)(\theta^\prime - \theta )} = (n_0 - i \partial_{\theta}) \langle \theta \vert \theta^\prime\rangle,
\end{equation}

which shows that the mapping to a derivative is exact.

\subsection{Floquet limit.} We restrict the oscillator Hilbert space to large photon numbers around $n_0\gg 0$, where the relevant values of $k \equiv n - n_0$ satisfy $|k|\ll n_0$,

\begin{equation}
	\ket{\theta} = \sum_{k= -n_0}^\infty e^{-i k \theta}\ket{n}.
\end{equation}

In the large-$n_0$ limit, the lower bound can be extended to $-\infty$, because the states contributing to the dynamics are localized near
$k=0$. The overlap then becomes the Dirac comb,

\begin{eqnarray}
	\langle \theta \vert \theta^\prime\rangle  =    \sum_{k= -n_0}^{\infty} e^{-ik(\theta^\prime - \theta)} \underbrace{\longrightarrow}_{n_0 \rightarrow \infty}   2\pi\delta_{2\pi}(\theta - \theta^\prime).
\end{eqnarray}

The matrix element $\langle \theta \vert \hat{a} \vert \theta^\prime \rangle$  can now be further simplified, by expanding

\begin{equation}
	\sqrt{n_0 + k} = \sqrt{n_0}
	\left[
	1+\frac{k}{2n_0}
	+\mathcal{O}\left(\frac{k^2}{n_0^2}\right)
	\right],
\end{equation}

and pulling out of the sum the corresponding factor in Eq. \eqref{eq:a_matrix_elements}, we arrive at

\begin{eqnarray}
	\langle \theta \vert \hat{a} \vert \theta^\prime \rangle & \approx & e^{-i\theta^\prime} \sqrt{n_0} \sum_{k = -n_0}^\infty  e^{-ik(\theta^\prime - \theta)} \nonumber \\
	& \underbrace{\longrightarrow}_{n_0\rightarrow \infty } &  2\pi e^{-i\theta^\prime} \,\sqrt{n_0}\, \,\delta_{2\pi}(\theta^\prime-\theta)
	\label{eq:floquetlimit_a}
\end{eqnarray}

Similarly, for the photon number operator we obtain

\begin{equation}
	\langle \theta \vert \hat{a}^{\dagger} \hat{a}|\theta^\prime \rangle
	\underbrace{\longrightarrow}_{n_0 \rightarrow \infty} 
	2\pi \left[ n_0  -  i\partial_\theta \right]\delta_{2\pi}(\theta - \theta^\prime). \label{eq:theta_numberphot}
\end{equation}

\subsection{New form of the eigenvalue equation} We can now write the dipole-gauge Hamiltonian in the new $\theta-$representation,

\begin{equation}
	\langle \theta \vert  \hat{H}_{\text{QRM}}^{(\text{D})} \vert \varphi_\beta \rangle = \left( \hat{H}_{\text{QRM}}^{(\text{D})} \vert \varphi_\beta \rangle \right)(\theta).
\end{equation}

Using the equalities in Eqs. \eqref{eq:floquetlimit_a} and \eqref{eq:theta_numberphot}, and operating the Dirac comb as in Eq. \eqref{eq:dirac_comb}, we obtain,

\begin{equation}
	\int_0^{2\pi} \frac{d\theta^\prime}{2\pi} \, \langle \theta \vert  \hat{H}_{\text{QRM}}^{(\text{D})}\vert \theta^\prime \rangle \langle \theta^\prime \vert \varphi_\beta \rangle = \hat{H}^{(\text{D})}_{\text{QRM}}(\theta) \ket{\varphi_\beta (\theta)},
\end{equation}

where

\begin{equation}
	\hat{H}^{(\text{D})}_{\text{QRM}}(\theta)  = \hat{H}_\text{mat} + 2g\sqrt{n_0} \sin(\theta) \cdot\hat{Z} + \Omega(n_0-i\partial_\theta).
\end{equation}

The self-dipole energy term in Eq.~\eqref{eq:quantum_dipole} scales as

\begin{equation}
	\frac{g^2}{\Omega}
	=
	\frac{A^2}{4\Omega n_0} ,
	\label{eq:quadratic_scaling}
\end{equation}

and therefore vanishes in the limit $n_0\to\infty$.

	\clearpage
	\onecolumngrid

	\appendix

	\makeatletter
	\@removefromreset{equation}{section}
	\@removefromreset{figure}{section}
	\@removefromreset{table}{section}
	\makeatother
	
	\setcounter{equation}{0}
	\setcounter{figure}{0}
	\setcounter{table}{0}
	
	\renewcommand{\thesection}{S\arabic{section}}
	\renewcommand{\theequation}{S\arabic{equation}}
	\renewcommand{\thefigure}{S\arabic{figure}}
	\renewcommand{\thetable}{S\arabic{table}}
	
	\vspace*{1em}
	
	\begin{center}
		\makebox[\textwidth]{
			\rule{0.12\textwidth}{0.5pt}
			\hspace{0.8em}
			{\large\bfseries Supplemental Material}
			\hspace{0.8em}
			\rule{0.12\textwidth}{0.5pt}
		}
		\vspace{0.5em}
		
		{\large\bfseries
			Floquet physics from quantized light-matter interaction: geometric phases, gauge
			consistency, and entanglement
		}
		
		\vspace{0.5em}
		{\normalsize Beatriz P\'erez-Gonz\'alez, Sigmund Kohler, M\'onica Benito
		}
		
	\end{center}
	
	\vspace{1.5em}

\title{\textit{Supplemental material to:} \\
	Floquet physics from quantized light-matter interaction: geometric phases, gauge consistency, and entanglement}
\author{Beatriz P\'erez Gonz\'alez}
\author{Sigmund Kohler}
\author{ M\'onica Benito}

\maketitle

\section{Anandan phase and mean energy \label{app:anandan_phase}}

\subsection{Driven systems \label{app:anandan_phase_floquet}}

Let us consider the averaged energy in a Floquet state $\ket{\Psi_\alpha(t)} = e^{-i\varepsilon_\alpha t}\ket{\Phi_\alpha (t)}$ \cite{GRIFONI1998229},

\begin{subequations}
	\begin{align}
		\overline{H}_\alpha & =  \frac{1}{T}\int^T_0 dt \bra{\Psi_\alpha(t) }H(t) \ket{\Psi_\alpha(t)} \label{eq:def_mean_energy_floquet} \\
		& =  \varepsilon_\alpha + \frac{1}{T}\int^T_0 dt \bra{\Phi_\alpha(t) } i\partial_t \ket{\Phi_\alpha(t)}.
	\end{align}
\end{subequations}

This connection between the averaged energy $\overline{H}_\alpha$ and the quasienergy $\varepsilon_\alpha$ in addition provides a relation between the dynamical and geometrical phase of the system. The dynamical phase is generally defined as the time-average expectation value of the Hamiltonian, i.e., $\overline{H}_\alpha$. After one period, the full state comes back to itself up to a phase, $\ket{\Psi_\alpha(T)} = e^{-i\varepsilon_\alpha T}\ket{\Psi_\alpha(0)}$, so the total phase after one period is $\chi=-\varepsilon_\alpha T$. Therefore, the geometric contribution is the remaining term \cite{MOORE19911, Moore_1990, PhysRevA.55.1653}, which is defined as the Anandan phase $\gamma_\alpha$

\begin{equation}
	\frac{\gamma_\alpha^{\rm F}}{T} \equiv  \overline{H}_\alpha   -\varepsilon_\alpha = \frac{1}{T}\int_0^T dt\langle \Phi_\alpha(t)\vert i\partial_t \vert \Phi_\alpha(t)\rangle .
	\label{eq:floquet_anandan_supp}
\end{equation}

\subsection{Quantum light \label{app:anandan_phase_quantum}}

To obtain the quantum counterparts of the Floquet mean energy and
Anandan phase, we focus on the correspondence between the photon-number operator and the phase
derivative $\hat{a}^\dagger\hat{a} \longrightarrow n_0-i\partial_\theta$, which implies

\begin{equation}
	n_0-
	\langle \hat{a}^\dagger\hat{a}\rangle_\beta
	=
	\frac{1}{2\pi}
	\int_0^{2\pi}
	d\theta\,
	\bra{\varphi_\beta(\theta)}
	i\partial_\theta
	\ket{\varphi_\beta(\theta)} .
	\label{eq:quantum_anandan_derivation}
\end{equation}

In direct analogy with the Floquet expression in
Eq.~\eqref{eq:floquet_anandan_supp}, we therefore define the quantum
Anandan phase as

\begin{equation}
	\frac{\gamma_\beta^{Q}}{2\pi}
	\equiv
	n_0-
	\langle \hat{a}^\dagger\hat{a}\rangle_\beta
	=
	\int_0^{2\pi}
	\frac{d\theta}{2\pi}\,
	\bra{\varphi_\beta(\theta)}
	i\partial_\theta
	\ket{\varphi_\beta(\theta)} .
	\label{eq:quantum_anandan}
\end{equation}

Equivalently, following Eq. \eqref{eq:def_mean_energy_floquet}, we can write for the quantum Hamiltonian

\begin{equation}
	\overline{E}_\beta
	=
	\frac{1}{2\pi}
	\int_0^{2\pi}
	d\theta\,
	\bra{\varphi_\beta(\theta)}
	\hat{h}^{(\mathrm{D})}_{\mathrm{QRM}}(\theta)
	\ket{\varphi_\beta(\theta)} .
\end{equation}

which amounts to subtracting the
average photonic energy from the total eigenenergy,

\begin{equation}
	\overline{E}_\beta
	\equiv
	E_\beta
	-
	\Omega
	\langle \hat{a}^\dagger\hat{a}\rangle_\beta .
	\label{eq:quantum_mean_energy_theta}
\end{equation}

Lastly, combining Eqs.~\eqref{eq:quantum_anandan} and
\eqref{eq:quantum_mean_energy_theta}, and using $T=2\pi/\Omega$, gives

\begin{equation}
	T\left(E_\beta-\Omega n_0\right)
	=
	T\overline{E}_\beta - \gamma_\beta^{Q}.
	\label{eq:quantum_phase_relation}
\end{equation}

This has exactly the same structure as the Floquet relation
$T\varepsilon_\alpha=T\overline{H}_\alpha -  \gamma_\alpha^{F}$.

\section{Numerical results for the semiclassical limit for different parameter regimes \label{app:disagreement}}

\begin{figure}
	\centering
	\includegraphics{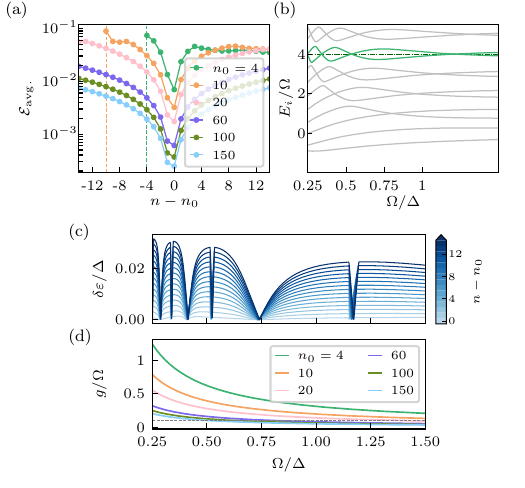}
	\caption{Validity of the quantum-Floquet correspondence for fixed $A / \Delta = 1.25$. (a) Frequency-averaged spectral mismatch $\mathcal{E}_{\mathrm{avg}}$ as a function of the photon-sector offset $n-n_0$ for the indicated $n_0$. The frequency range considered is that of Fig. \ref{fig:quasi_meane_anadan}. (b) Quantum spectrum for $n_0=4$, highlighting the photon sector around the energy $n_0\Omega$. The asymmetry in the upper and lower branches account for the asymmetry shown by $\mathcal{E}_{\mathrm{avg.}}$ in panel (a). (c) Frequency-resolved mismatch $\Omega\mathcal{E}$ for $n_0=150$ and different offsets $n-n_0$. (d) Effective normalized light-matter coupling $g/\Omega$ for the same values of $n_0$. Horizontal line marks $g/\Omega = 0.1$.    \label{fig:further_comparison}}
\end{figure}

In this section, we examine how the quantum description approaches the semiclassical Floquet limit for different parameter choices ($n_0, A/\Delta, \Omega / \Delta$). We first provide additional numerical details supporting Fig.~\ref{fig:quasi_meane_anadan}. Figure~\ref{fig:further_comparison}(a) quantifies the agreement between the Floquet mode with quasienergy $\varepsilon_0$, and the corresponding quantum state $E_i(n)$ centered around $n\Omega$, where $(n-n_0)$ labels the photon sector above $(n-n_0>0)$ or below $(n-n_0<0)$ the reference sector $n_0$. The spectral mismatched $\delta \varepsilon = \vert \varepsilon_0 - (E_i(n) - n\Omega)  \vert$ is then averaged over the entire frequency range, yielding $\mathcal{E}_{\mathrm{avg.}} = \langle \delta \varepsilon /\Omega \rangle_{\Omega}$. The mismatch is smallest around $n=n_0$, and increases as one moves towards more distant photon sectors. Increasing $n_0$ substantially enlarges the range of sectors that reproduce the Floquet spectrum with a rather small error. The vertical dashed lines for $n_0=4$ and $n_0 = 12$ indicate the lower boundary $n=0$, beyond which no physical photon sectors exist. The asymmetry between $(n - n_0) >0$ and $(n - n_0) < 0$, specially for the lower values of $n_0$ can be explained by the asymmetry between the two states forming the doubled associated with the $n$ photonic band. This is illustrated directly in Fig.~\ref{fig:further_comparison}(b) for $n_0 = 4$. Thus, a large photon occupation is required for agreement over a broad energy window, but not necessarily for reproducing a restricted set of Floquet states.\\
Figure~\ref{fig:further_comparison}(c) resolves the mismatch $ \delta \mathcal{\varepsilon} / \Delta$ as a function of $\Omega/\Delta$ for $n_0=150$. The agreement persists throughout the considered frequency range for sectors close to $n_0$, while deviations progressively increase with $(n-n_0)$. Finally, Fig.~\ref{fig:further_comparison}(d) shows the bare normalized coupling $\frac{g}{\Omega}
= A / 2\sqrt{n_0}\Omega$, obtained while keeping the effective classical driving amplitude $A$ fixed. As $n_0$ increases, $g/\Omega$ decreases and the quantum spectrum converges over an increasingly broad range of photon sectors. \\

\begin{figure}
	\centering
	\includegraphics{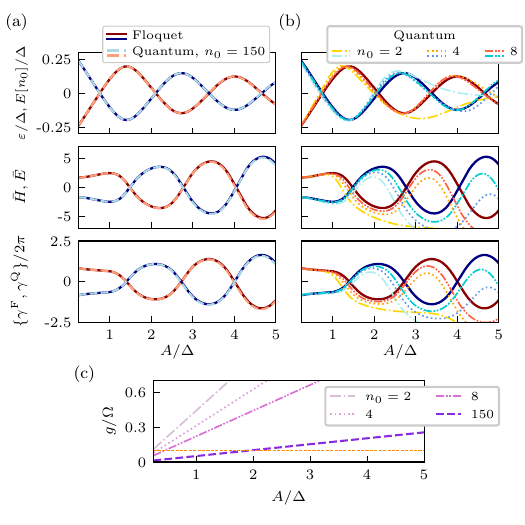}
	\caption{Quantum-Floquet correspondence versus driving amplitude. Quasienergies, mean energies, and Anandan phases are compared for (a) $n_0=150$ and (b) $n_0=2,4,8$, with the Floquet result shown by solid curves and the quantum results depicted through discontinuous lines. The parameter $\Omega, \, \omega = 0.8\Delta$ is fixed. (c) Corresponding normalized light-matter coupling $g/\Omega$; the horizontal line marks $g/\Omega=0.1$.}
	\label{fig:comparison_A}
\end{figure}

A complementary view of the convergence is provided in Fig.~\ref{fig:comparison_A}, where the frequency is fixed to $\Omega = 0.8\Delta$ and the amplitude $A/\Delta$ is varied. Note that we are exploring a low-frequency regime, in which the disagreement for lower $n_0$ is stronger for a given $A$. For $n_0=150$ [Fig.~\ref{fig:comparison_A}(a)], the quantum quasienergies, mean energies, and Anandan phases remain in close agreement with their Floquet counterparts throughout the considered amplitude range, as expected. By contrast, for the lower occupations \(n_0=2,4,\) and \(8\) [Fig.~\ref{fig:comparison_A}(b)], the correspondence progressively deteriorates as $A/\Delta$ increases. Notably, $n_0 = 2$ already gives a good correspondence for the Floquet result for small driving amplitudes ($A/\Delta \lesssim 1$). This highlights the interplay of all the parameter choices when evaluating the validity of the quantum-to-classical crossover. Panel~\ref{fig:comparison_A}(c) shows the effective light-matter coupling obtained from each $n_0$ choice. The horizontal line at \(g/\Omega=0.1\) is included as a visual guide. The ratio $g/\Omega$ gets larger than $\sim 0.1$ even for $n_0 = 150$ for sufficiently large $A/\Delta$, while still preserving the quantum and semiclassical agreement.  These results show that convergence towards the Floquet limit is governed not only by the photon occupation \(n_0\), but by the combined scaling of \(n_0\), \(A\), and \(\Omega\).

\section{ Gauge transformation \label{app:gauge}}

\subsection{Gauge transformation in the Quantum Rabi model \label{app:change_gauge}}

Here we derive the dipole- and Coulomb-gauge Hamiltonians used in the
main text following the projected-unitary construction of
Ref.~\cite{Olesya2021}. We consider the two-level matter system and the 
single photonic mode $\hat H_{\mathrm{mat}}=\frac{\Delta}{2}\hat\sigma_x$, and $\hat H_{\mathrm{ph}}=\Omega \hat a^\dagger \hat a$, with $\hat{Z} = \hat\sigma_z$. Gauge-equivalent Hamiltonians can be generated from the projected unitary transformation

\begin{equation}
	\hat U_{\mathrm{gauge}}
	= \exp\left[ i(\hat{a} + \hat{a}^\dagger )\sum_{i,j} \hat{c}^\dagger_i  \, \chi_{i,j} \, \hat{c}_j\right]
	\label{eq:gauge_unitary}
\end{equation}

where $\hat{c}^{(\dagger)}_i$ are the fermionic creation/annihilation operators, and the gauge function associated with the cavity vector potential $\nabla \, \chi(\mathbf{r}) = e A(\mathbf{r})$ gives the matrix elements of $\chi(\mathrm{r})$ in the Wannier basis, $\chi_{ij} = \int d\mathbf{r} \langle i\vert \chi(\mathbf{r})\vert j \rangle$. Under the long-wavelength approximation in our 1D system, $A_0(\mathbf{r}) \approx A_0$, and $\chi_{i,j}$ depends directly on the electric-dipole matrix elements

\begin{equation}
	\chi_{i,j} = eA_0\langle i \vert \hat{x} \vert j\rangle = eA_0 x_i \delta_{i,j}.
\end{equation}

We can interpret our TLS as describing a two-site system, $\ket{L}$ (left) and $\ket{R}$ (right). With this, 

\begin{equation}
	\sum_{i,j} \hat{c}^\dagger_i  \, \chi_{i,j} \, \hat{c}_j = x_L\hat{c}^\dagger_L \hat{c}_L + x_R\hat{c}^\dagger_R \hat{c}_R = (d/2) \,\hat{\sigma}_z
\end{equation}

where $d = \vert x_L-x_R\vert$, $x_L = - d/2$ and $x_R = d/2$, and $\hat{\sigma}_z = \vert R\rangle \langle R\vert  - \vert L \rangle \langle L \vert$. In the following, we define $\zeta \equiv e A_0 d/2$. Finally, the unitary transformation becomes

\begin{equation}
	\hat U_{\mathrm{gauge}} =  \exp\left[ i\zeta(\hat{a} + \hat{a}^\dagger ) \hat{\sigma}_z \right].
\end{equation}

The dipole-gauge Hamiltonian is obtained by dressing the photonic Hamiltonian while leaving the bare matter Hamiltonian unchanged,

\begin{equation}
	\hat H^{(\mathrm{D})}_{\mathrm{QRM}}
	=
	\hat H_{\mathrm{mat}}
	+
	\hat U_{\mathrm{gauge}} \, \hat H_{\mathrm{ph}} \, \hat U^\dagger_{\mathrm{gauge}}.
	\label{eq:HD_general}
\end{equation}

Using

\begin{subequations}
	\begin{align}
		\hat U_{\mathrm{gauge}}\,  \hat a \, \hat U^\dagger_{\mathrm{gauge}}
		& =
		\hat a -i\zeta\,\hat\sigma_z, \\
		\hat U_{\mathrm{gauge}} \,  \hat a^\dagger \, \hat U^\dagger_{\mathrm{gauge}}
		& =
		\hat a^\dagger+i\zeta \,\hat\sigma_z,
	\end{align}
\end{subequations}

we obtain

\begin{equation}
	\hat{H}^{\mathrm{(D)}}_{\mathrm{QRM}} =
	\frac{\Delta}{2}\hat\sigma_x
	+ \Omega \hat a^\dagger\hat a
	+i\Omega \zeta(\hat{a} - \hat{a}^\dagger)\hat{\sigma}_z + \zeta^2\Omega \, \hat{\sigma}_z^2
\end{equation}

The last term, as shown in this calculation, is required by the gauge transformation. For a two-level system with
dipole operator $\hat\sigma_z$, it is proportional to the identity
and therefore only produces a global energy shift. Now, by defining $g \equiv \zeta \Omega$, we arrive

\begin{equation}
	\hat{H}^{\mathrm{(D)}}_{\mathrm{QRM}} =
	\frac{\Delta}{2}\hat\sigma_x
	+ \Omega \hat a^\dagger\hat a
	+ig(\hat{a} - \hat{a}^\dagger)\hat{\sigma}_z + \frac{g^2}{\Omega}
\end{equation}

which is the dipole-gauge Hamiltonian as shown in Eq. \eqref{eq:quantum_dipole}.\\

The gauge-equivalent Coulomb-gauge Hamiltonian is instead obtained by
dressing the matter Hamiltonian,

\begin{equation}
	\hat{H}_{\mathrm{QRM}}^{(\mathrm{C})}
	=
	\hat H_{\mathrm{ph}}
	+
	\hat U^\dagger_\mathrm{gauge}\,
	\hat H_{\mathrm{mat}}\,
	\hat U_\mathrm{gauge}.
	\label{eq:HC_general}
\end{equation}

which yields

\begin{equation}
	\hat{H}_{\text{QRM}}^{(\text{C})} = \Omega \hat{a}^\dagger \hat{a} \, + \,\frac{\Delta}{2}\left[ e^{-i2\zeta(\hat{a} + \hat{a}^\dagger)} \, \hat{\sigma}_+ + \text{h.c.}\right].
	\label{eq:coulom_sm}
\end{equation}

With $\eta = 2\zeta = 2g/\Omega$, one can recover the expression of Eq. \eqref{eq:QRM_Coulomb}.\\

The equivalence between the two representations follows directly from

\begin{equation}
	\hat{H}^{\mathrm{(D)}}_{\mathrm{QRM}}
	=
	\hat U_\mathrm{gauge} \,\hat{H}_{\text{QRM}}^{(\text{C})} \, \hat U^\dagger_\mathrm{gauge}.
	\label{eq:gauge_equivalence}
\end{equation}

Consequently, both Hamiltonians possess exactly the same energy
spectrum when the full nonlinear Coulomb-gauge Hamiltonian structure is
retained.\\

\subsection{Frame transformation for driven systems \label{app:floquet_frame}}

Here we derive the Coulomb-frame semiclassical Hamiltonian used in the main text. Starting from the dipole-frame Hamiltonian, we apply the time-dependent unitary transformation

\begin{align}
	\hat{U}_{\mathrm{driv}}(t) & =
	\exp\left[ -iA\hat{\sigma}_z\int^t dt'\,\sin(\omega t') \right] \nonumber \\
	& = \exp\left[ i \, \frac{A}{\omega} \,\cos(\omega t)\hat{\sigma}_z
	\right].
	\label{eq:semiclassical_frame_unitary}
\end{align}

The transformed Hamiltonian is

\begin{align}
	\hat{H}^{(\mathrm{C})}_{\mathrm{SCRM}}(t) & =
	\hat{U}_{\mathrm{driv}}^\dagger(t)
	\hat{H}^{(\mathrm{D})}_{\mathrm{SCRM}}(t)
	\hat{U}_{\mathrm{driv}}(t) \nonumber \\
	& \quad \quad \quad \quad \quad \quad -
	i\hat{U}_{\mathrm{driv}}^\dagger(t)
	\partial_t\hat{U}_{\mathrm{driv}}(t).
\end{align}

Writing $\hat{\sigma}_x=\hat{\sigma}_+ + \hat{\sigma}_-$ and using

\begin{equation}
	\hat{U}_{\mathrm{driv}}^\dagger(t)
	\hat{\sigma}_{\pm}
	\hat{U}_{\mathrm{driv}}(t)
	=
	e^{\mp i(2A/\omega)\cos(\omega t)}
	\hat{\sigma}_{\pm},
\end{equation}

we obtain

\begin{equation}
	\hat{H}^{(\mathrm{C})}_{\mathrm{SCRM}}(t)
	=
	\frac{\Delta}{2}
	\left[
	e^{-i\frac{2A}{\omega}\cos(\omega t)}
	\hat{\sigma}_{+}
	+ \text{h.c.}
	\right].
	\label{eq:semiclassical_coulomb_frame}
\end{equation}

The frame transformation therefore removes the longitudinal drive and transfers its time dependence to a phase dressing of the tunneling term. This is the semiclassical counterpart of the Coulomb-gauge Hamiltonian obtained from the fully quantized model.

\section{Floquet limit for non-gauge-equivalent Hamiltonians \label{app:non-gi-models}}

Expanding Eq.~\eqref{eq:coulom_sm} for $g/\Omega\ll1$ gives the linearized version of the Coulomb-gauge quantum Hamiltonian,

\begin{equation}
	\hat{H}^{(\text{C})}_{\text{ngi-QRM}} = \Omega \hat{a}^\dagger \hat{a} +\frac{\Delta}{2}\hat{\sigma}_x + g_C (\hat{a}^\dagger + \hat{a})\hat{\sigma}_y 
	\label{eq:non-gaugeinv-coulomb}
\end{equation}

where the diamagnetic contribution has been discarded, as is customary in the small-coupling regime. The effective coupling $g_C = \Delta \eta/2 = \Delta g/\Omega$ now depends explicitly on the tunneling amplitude $\Delta$. The required semiclassical limit should read $2g_{\rm C}\sqrt{n_0} = A$. The distinction between $g$ and $g_C$ is essential, especially away from resonance $\Delta = \Omega$. More generally, gauge equivalence requires retaining the complete trigonometric dependence in Eq.~\eqref{eq:coulom_sm}.\\

\begin{figure}
	\centering
	\includegraphics{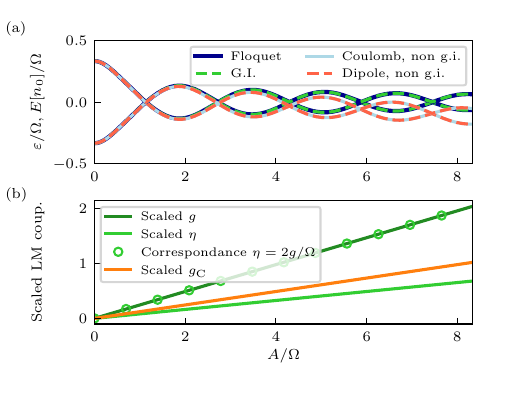}
	\caption{(a) Floquet quasienergies compared with the gauge-invariant (G.I.) quantum model and the linearized non-gauge-invariant Coulomb- and dipole-gauge models. For all, $n_0 = 150$, and $\Omega = 1.5\Delta$. (b) Corresponding scaled light--matter couplings $g$, $\eta$, and $g_C$, including the relation $\eta = 2g/\Omega$.}
	\label{fig:non_gi}
\end{figure}

To the same order in the light-matter coupling, the dipole gauge Hamiltonian gives

\begin{equation}
	\hat{H}^{(\text{D})}_{\text{ngi-QRM}} = \Omega \hat{a}^\dagger \hat{a} + \frac{\Delta}{2}\hat{\sigma}_x + ig \left(\hat{a} - \hat{a}^\dagger \right) \hat{\sigma}_z,
	\label{eq:non-gaugeinv-dipole}
\end{equation}

with the corresponding semiclassical scaling $2g\sqrt{n_0} \equiv A$. Notice that the dipole self-energy term is omitted. 
The inconsistency of the linearized quantum models becomes apparent when their spectra are compared with the corresponding Floquet quasienergies, as shown in Fig.~\ref{fig:non_gi}(a). The gauge-invariant models used in the main text (dashed green) reproduce the Floquet quasienergies (solid blue) over the full range of $A/\Omega$, whereas the Coulomb- and dipole-gauge models of Eqs.~\eqref{eq:non-gaugeinv-coulomb} and \eqref{eq:non-gaugeinv-dipole}, respectively, progressively deviate at large driving amplitudes. Figure~\ref{fig:non_gi}(b) highlights the origin of this discrepancy by comparing the different definitions of the light--matter coupling strength \cite{PhysRevA.98.053819}. For the gauge-equivalent Hamiltonians, the dipole- and Coulomb-gauge couplings are consistently related by $\eta=2g/\Omega$, and the same semiclassical scaling can therefore be imposed in either gauge. By contrast, Eqs.~\eqref{eq:non-gaugeinv-coulomb} and \eqref{eq:non-gaugeinv-dipole} are connected by the gauge transformation
$\hat U_{\rm ngi} = \exp\!\left[ i\frac{g}{\Omega}
(\hat a+\hat a^\dagger)\hat\sigma_z
\right]$ only to linear order in $g/\Omega$. Their gauge equivalence is consequently lost at higher orders, since the terms generated by the full unitary transformation have been truncated. At resonance, $\Delta=\Omega$, one additionally has $g_{\rm C}=g$, and the two linearized Hamiltonians can be mapped exactly onto one another through a $\pi/2$ rotation in oscillator phase space combined with a spin rotation. This special equivalence is a basis transformation rather than the full gauge transformation and does not persist away from resonance.

\section{Photon statistics in the dipole gauge}

\begin{figure}
	\centering
	\includegraphics{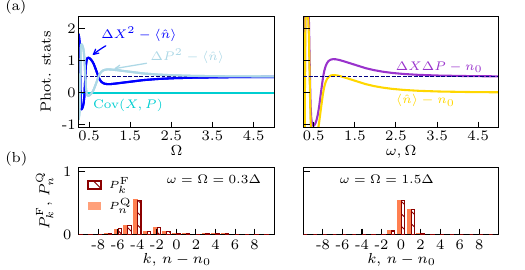}
	\caption{(a) Dipole-gauge photon-statistics indicators versus $A/\Omega$, for fixed $\Omega= 0.6\Delta$: $\Delta X^2-\langle\hat n\rangle$, $\Delta P^2-\langle\hat n\rangle$, $\mathrm{Cov}(X,P)$, $\Delta X\,\Delta P$, and $\langle\hat n\rangle-n_0$. (b) Quantum photon-number distribution \(P_n^{\mathrm Q}\) and Floquet harmonic weights \(P_k^{\mathrm F}\) at \(A/\Omega=1\) and \(5\), respectively. For all plots, $n_0 = 150$.}
	\label{fig:phot_stats_dip}
\end{figure}

We now compare the photonic statistics obtained in the Coulomb and dipole gauges. The latter are shown in Fig. \ref{fig:phot_stats_dip}. In both cases, the quantum photon-sector weights can be directly compared with the Fourier weights of the corresponding Floquet mode, supporting the identification between photon sectors and Sambe harmonics in the Floquet limit. At weak driving, the two gauges give qualitatively similar results: the photonic distribution remains localized around the reference sector $n_0$, the covariance $\mathrm{Cov}(X,P)$ is negligible, and the quadrature variances reduce to the Fock-state value $\Delta X^2=\Delta P^2=\langle \hat n\rangle+1/2$ at $A=0$.
The behavior at larger driving amplitudes, however, is markedly gauge dependent. In the Coulomb gauge, increasing $A$ mainly produces nearly equal-and-opposite oscillations of $\Delta X^2-\langle \hat n\rangle$ and $\Delta P^2-\langle \hat n\rangle$, while the photon-number distribution remains comparatively narrow and centered close to $n_0$. The uncertainty product $\Delta X\Delta P$ follows the oscillations of $\langle \hat n\rangle$, but both quantities remain moderately displaced from their $A=0$ values. By contrast, in the dipole gauge the same increase in $A$ produces a much stronger redistribution of photonic weight over neighboring Fock sectors. This is reflected in larger oscillations of both $\langle \hat n\rangle-n_0$ and $\Delta X\Delta P-n_0$, as well as in a broader and more structured photon-number distribution at large $A/\Omega$. These differences reflect the fact that the notion of photonic content depends on the gauge used to partition the total light-matter system into matter and field degrees of freedom. The comparison therefore illustrates an important point: the quantum-to-classical mapping is gauge consistent, but the apparent photonic statistics associated with a given Floquet mode are not universal. They encode how the same physical state is represented in a particular gauge.
	
\end{document}